\documentclass{My_WileyChemistry-template}
\usepackage{hyperref}
\usepackage{cleveref}
\usepackage{cuted}
\DeclareGraphicsExtensions{.jpg, .eps, .ps, .pdf}
\graphicspath{{../data/}{.}}

\title{\raggedright {Unlocking n-alk-1-ynes Conformers: Quantum ``Trigger Finger'' versus
    ``Stiff Joint'' Conformations}}

\author{
\begin{minipage}{\textwidth}
	Ioan B\^aldea \textsuperscript{[a]}
\end{minipage}
}

\newcommand{\affiliation}{
\begin{itemize}

\item[{[a]}] Dr. Ioan B\^aldea\\
Theoretical Chemistry, Heidelberg University, Im Neuenheimer Feld 229, D-69120 Heidelberg, Germany\\
E-mail: ioan.baldea@pci.uni-heidelberg.de
\end{itemize}
}

\renewcommand{\abstract}{
  Molecular conformation in n-alk-1-ynes (CnA) is conventionally simplified to an all-planar structure.
  We report a comprehensive quantum chemical analysis revealing two near-isoenergetic rotamers at the acetylenic terminus: planar (C$_s$) and skewed (C$_1$).
  The high, symmetric rotational energy barrier ($\approx 150$\,meV) arises from unique steric relief near the $\mathrm{sp}$ center coupled with electronic stabilization of C$_1$. This creates a unique kinetic profile: a Quantum ``Trigger Finger'' ($\alpha$ rotation) that enforces an $\approx 50\%:\,50\%$ $\mathrm{C}_s/\mathrm{C}_1$ ensemble, sharply contrasting with the thermodynamically biased ``Stiff Joint'' ($\delta$ rotation) of the alkyl chain. This structural degeneracy necessitates ensemble averaging for spectroscopic data interpretation, while the slow interconversion permits kinetic trapping and intentional conformer enrichment during synthesis and molecular junction fabrication. Our work redefines the alkyne anchor,
  providing a blueprint for accurate interpretation of spectroscopic data and achieving conformational control in molecular electronics.\\
}

\newcommand{\keywords}{
	Alkyne Conformation\textbullet\
	Hyperconjugation \textbullet\
	Quantum Chemistry \textbullet\
	Kinetic Control \textbullet\
	Molecular Electronics
}

\begin{document}

\twocolumn[\vspace{-1.5cm}\maketitle\vspace{-1cm}]
\small{\begin{shaded}
		\noindent\abstract
	\end{shaded}
}

\begin{figure} [!b]
\begin{minipage}[t]{\columnwidth}{\rule{\columnwidth}{1pt}\footnotesize{\textsf{\affiliation}}}\end{minipage}
\end{figure}

\section*{Introduction} \label{sec:intro}

While alkyne ($\ce{C\bond{3}CH}$) terminals possess favorable properties (stability and strong electronic coupling) that suggest their broad use as robust anchors for molecular devices, this potential has not translated into widespread adoption. In the field of molecular junctions,
thiols overwhelmingly dominate the literature,\cite{Frisbie:01,Rampi:01,Wang:03,York:03,Tao:03,Haiss:04,Tao:06c,Reddy:06,Wandlowski:08c,Guo:11,Zandvliet:12,Baldea:2019a,Baldea:2019h}
and the use of n-alk-1-ynes (CnA, \ce{HC\equiv C-CF2-(CH2)_{n-1}-CH3})
in fabricating such devices is exceedingly rare, with only a few publications \cite{Chiechi:14,Whitesides:15,Baldea:2024d,Baldea:2026a}
reporting CnA-based junctions to date.

This minimal experimental adoption is coupled with a persistent theoretical blind spot: the prevailing convention assumes an all-planar alkyne conformation (C$_s$), a fact reinforced by the total absence of the non-planar C$_1$ conformer (to which the present study is mainly devoted) in the NIST database.\cite{webbook}

Importantly in this context,  experimental spectroscopic studies (IR, Raman, microwave) of CnA conformers in the gas phase
  revealing (meta)stable non-planar conformers
  are limited to very short molecules: 1-butyne (C2A), \cite{Crowder:86} 1-pen\-tyne (C3A), \cite{Crowder:86,Crowder:88,Holme:09} and 1-hexyne (C4A). \cite{Crowder:88,Atticks:01,Utzat:07}
  These mo\-le\-cules are too short to function as molecular junctions; short-circuiting issues preclude their use in fabricating robust devices.
  Furthermore, while those studies identified the presence of non-planar conformers,
  they did not address the fundamental mechanistic question of \textit{whether or why} the non-planar form is stable.
  The crucial interplay of steric relief at the $\mathrm{sp}$-hybridized carbon and electronic stabilization via
  $\pi_{\ce{C\bond{3}C}} \rightarrow \sigma^{\ast}_{\ce{C-H}}$ hyperconjugation---an essential aspect to be discussed below---remained unexplored.

Recent transport measurements have revealed dual conductance regimes in CnA-based junctions,\cite{Baldea:2026a} hinting at underlying conformational complexity. However, a comprehensive quantum chemical understanding of the al\-kyne anchor's conformational landscape remains lacking.
This fundamental gap in both experimental data and foundational structural
recognition—which our investigation was triggered by—necessitates
a comprehensive quantum chemical re-eva\-lua\-tion of the conformational landscape of the alkyne anchor.

Our work presents a comprehensive quantum chemical analysis that resolves this long-standing ambiguity, revealing that the acetylenic terminus acts as a unique conformational element with two stable rotamers (C$_s$ and C$_1$).
The findings reported below demonstrate that the C$_1$ conformer is an intrinsic,
kinetically persistent component of the molecular conformer ensemble, a fact vital, e.g., for accurate interpretation of spectroscopic data and measurements on molecular electronic devices.
First made public as a preprint \cite{baldea_rotamers_CnA_chemrxiv} (cited as ref.~8 in \cite{Baldea:2026a}),
the theoretically predicted C$_s$/C$_1$ near-degeneracy---on which the present paper elaborates---
preceded, guided, and was validated by the joint theory–experiment investigation \cite{Baldea:2026a} carried out recently
in collaboration with an experimental group.

The persistent coexistence of C\textsubscript{s} and C\textsubscript{1} rotamers, locked by a high kinetic barrier, endows CnA molecules with a form of conformational memory, positioning them as \textit{smart molecular anchors} whose interface properties can be controlled through kinetic manipulation.

\section*{Results and Discussion}

\subsection*{Conformational Profile: The Terminal $\alpha$ Dihedral}

\begin{figure*}
\centerline{\includegraphics[width=0.9\textwidth,angle=0]{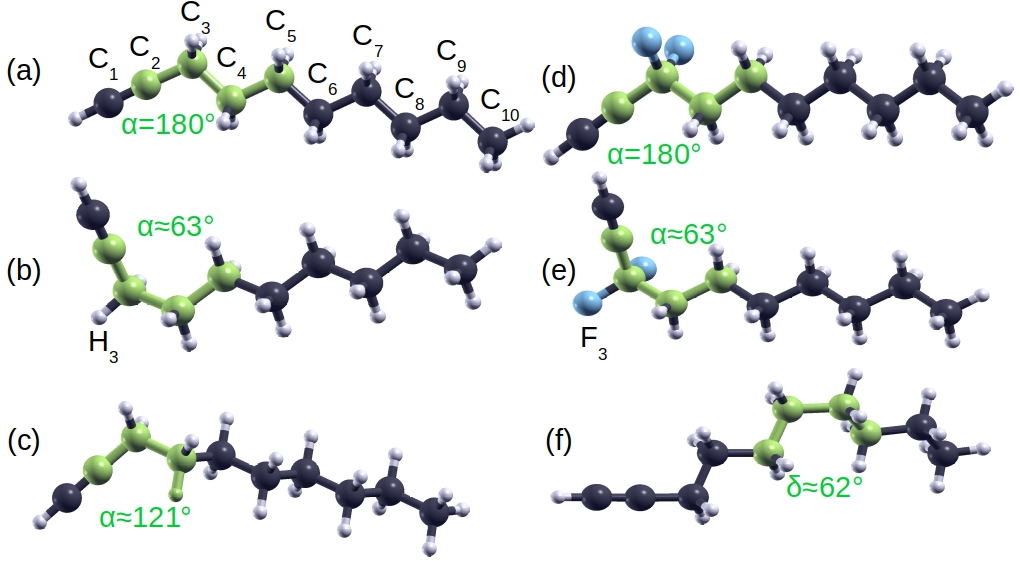}}
\caption{Optimized geometries for 1-decyne (C8A) and fluorinated decyne ($\mathrm{F2-C8A}$). The C atoms defining the $\alpha$- and $\delta$-dihedral angles are highlighted in green. (a, b) Planar (C$_s$) and skewed (C$_1$) $\alpha$-conformers of C8A. (c) Eclipsed transition state (trans-TS, $\alpha \approx 120^\circ$). (d, e) Planar (C$_s$) and skewed (C$_1$) $\alpha$-conformers of $\mathrm{F2-C8A}$. (f) Non-planar conformer with internal $\mathrm{gauche}$ motif ($\delta \approx 62^\circ$). IUPAC numbering is shown in panel (a).}
\label{fig:geometries}
\end{figure*}

Our investigation confirms the existence of two stable, low-energy rotamers governed by the terminal $\alpha$ dihedral angle: the planar (C$_s$, $\alpha = 180^\circ$) and the non-planar skewed (C$_1$, $\alpha \approx 63^\circ$) conformers
(\cref{fig:geometries}, bond metric data in
\cref{tab:metrics-planar-skewed-M062X-yes}).
\begin{table*}[!h]
\centering
\caption{Bond metrics versus size for planar and skewed CnA = \ce{H-C\bond{3}C-(CH2)_{n-1}-CH3}
  conformers optimized using the M06-2X exchange-correlation functional with GD3 dispersion corrections and Dunning's cc-pVTZ basis sets.\cite{g16}}
\label{tab:metrics-planar-skewed-M062X-yes}
\small
\begin{tabular}{rrrrrrrrrr}
\toprule
 CnA & Conformer & max r(H,H) & r(C1,Cn) &  r(C1,C2) &  r(C2,C3) &  r(C3,C4) & $ \angle\ce{C2C3C4C5}$  & $ \angle \ce{C2C3C4} $ & $ \angle \ce{C3C4C5} $ \\
\midrule
 C2A & planar & 5.5488 & 3.5208 & 1.1975 & 1.4609 & 1.5302 & N/A & 112.208 & N/A \\
     & skewed & 5.5487 & 3.5206 & 1.1975 & 1.4609 & 1.5302 & N/A & 112.197 & N/A \\
 C3A & planar & 6.6685 & 4.9566 & 1.1975 & 1.4599 & 1.5320 & 179.999 & 112.670 & 111.548 \\
     & skewed & 5.5460 & 3.8170 & 1.1977 & 1.4613 & 1.5341 & 62.712 & 112.571 & 112.552 \\
 C4A & planar & 7.9901 & 6.0336 & 1.1975 & 1.4599 & 1.5313 & 180.000 & 112.647 & 112.152 \\
     & skewed & 6.7599 & 5.1565 & 1.1977 & 1.4613 & 1.5332 & 63.344 & 112.642 & 113.133 \\
 C5A & planar & 9.1685 & 7.4223 & 1.1975 & 1.4599 & 1.5315 & 179.997 & 112.643 & 112.121 \\
     & skewed & 7.5016 & 5.9347 & 1.1977 & 1.4613 & 1.5334 & 63.369 & 112.623 & 113.105 \\
 C6A & planar & 10.4827 & 8.5677 & 1.1975 & 1.4599 & 1.5314 & 180.000 & 112.647 & 112.116 \\
     & skewed & 8.5622 & 7.3203 & 1.1977 & 1.4614 & 1.5333 & 63.355 & 112.622 & 113.084 \\
 C7A & planar & 11.6875 & 9.9286 & 1.1975 & 1.4598 & 1.5314 & 179.993 & 112.649 & 112.114 \\
     & skewed & 9.7357 & 8.2885 & 1.1977 & 1.4613 & 1.5333 & 63.369 & 112.619 & 113.083 \\
 C8A & planar & 12.9954 & 11.1065 & 1.1975 & 1.4599 & 1.5314 & 179.997 & 112.651 & 112.113 \\
     & skewed & 10.8224 & 9.6720 & 1.1977 & 1.4613 & 1.5333 & 63.255 & 112.629 & 113.084 \\
 C9A & planar & 14.2145 & 12.4497 & 1.1975 & 1.4599 & 1.5314 & 179.994 & 112.654 & 112.113 \\
     & skewed & 12.0938 & 10.7295 & 1.1977 & 1.4613 & 1.5333 & 63.323 & 112.624 & 113.090 \\
C10A & planar & 15.5185 & 13.6474 & 1.1975 & 1.4599 & 1.5314 & 179.999 & 112.654 & 112.110 \\
     & skewed & 13.2306 & 12.1002 & 1.1977 & 1.4614 & 1.5333 & 63.382 & 112.618 & 113.085 \\
C11A & planar & 16.7474 & 14.9789 & 1.1975 & 1.4599 & 1.5314 & 179.994 & 112.657 & 112.110 \\
     & skewed & 14.5186 & 13.2105 & 1.1978 & 1.4613 & 1.5333 & 63.101 & 112.645 & 113.098 \\
C12A & planar & 18.0478 & 16.1896 & 1.1975 & 1.4599 & 1.5314 & 179.996 & 112.656 & 112.110 \\
     & skewed & 15.7648 & 14.5735 & 1.1977 & 1.4614 & 1.5333 & 63.133 & 112.640 & 113.097 \\
C13A & planar & 19.2832 & 17.5125 & 1.1975 & 1.4599 & 1.5314 & 179.997 & 112.657 & 112.109 \\
     & skewed & 17.0318 & 15.7096 & 1.1978 & 1.4614 & 1.5333 & 63.217 & 112.635 & 113.094 \\
C14A & planar & 20.5803 & 18.7320 & 1.1975 & 1.4599 & 1.5314 & 179.999 & 112.655 & 112.109 \\
     & skewed & 18.3081 & 17.0618 & 1.1977 & 1.4614 & 1.5333 & 63.244 & 112.629 & 113.095 \\
C15A & planar & 21.8211 & 20.0488 & 1.1975 & 1.4598 & 1.5315 & 179.991 & 112.660 & 112.107 \\
     & skewed & 19.5758 & 18.2186 & 1.1977 & 1.4614 & 1.5333 & 63.381 & 112.615 & 113.089 \\
\bottomrule
\end{tabular}
\end{table*}

Multiple DFT and high-level composite thermochemistry methods consistently confirmed that C$_s$ and C$_1$ are
nearly isoenergetic, with $\Delta G < 0.2\,\mathrm{kcal/mol}$
(\cref{tab:conformer_stability_gibbs}).

\begin{table*}[htbp]
\centering
\caption{Conformer stability difference ($\Delta G = G_{\text{skewed}} - G_{\text{planar}}$) based on Gibbs Free Energy in kcal/mol (\text{298 K}) for $n$-alk-1-yne molecules (CnA) spanning C2A to C15A.
  A positive sign indicates the planar conformer is more stable, while a negative sign indicates the skewed conformer is more stable, as confirmed by the indicator in parentheses.
  The minor sign changes between methods are expected for such a near-degenerate system ($\vert\Delta G\vert < 0.2$ kcal/mol) and are within the intrinsic error limits of the composite methods. The G4 and CBS-QB3 methods, which include more complete basis set extrapolations and higher-order correlation treatments, are generally considered the most reliable for these subtle energetics and consistently show the skewed conformer to be slightly favored for most chain lengths.
}
\label{tab:conformer_stability_gibbs}
\begin{tabular}{r c c c c}
\toprule
\textbf{CnA} & \textbf{G3} & \textbf{G4} & \textbf{CBS-QB3} & \textbf{CBS-4M} \\
\midrule
C2A & -0.002 (skewed) & 0.000 (equal) & +0.001 (planar) & -0.006 (skewed) \\
C3A & -0.044 (skewed) & +0.034 (planar) & -0.009 (skewed) & -0.028 (skewed) \\
C4A & -0.061 (skewed) & +0.044 (planar) & -0.011 (skewed) & -0.068 (skewed) \\
C5A & -0.090 (skewed) & +0.029 (planar) & -0.011 (skewed) & -0.116 (skewed) \\
C6A & -0.100 (skewed) & +0.015 (planar) & -0.024 (skewed) & -0.149 (skewed) \\
C7A & -0.122 (skewed) & -0.005 (skewed) & -0.017 (skewed) & -0.142 (skewed) \\
C8A & -0.073 (skewed) & -0.033 (skewed) & -0.068 (skewed) & -0.158 (skewed) \\
C9A & -0.090 (skewed) & -0.028 (skewed) & -0.035 (skewed) & -0.142 (skewed) \\
C10A & -0.118 (skewed) & -0.048 (skewed) & -0.083 (skewed) & -0.164 (skewed) \\
C11A & -0.100 (skewed) & +0.057 (planar) & +0.034 (planar) & -0.154 (skewed) \\
C12A & -0.081 (skewed) & +0.063 (planar) & -0.018 (skewed) & -0.161 (skewed) \\
C13A & -0.109 (skewed) & 0.000 (equal) & 0.000 (equal) & -0.157 (skewed) \\
C14A & -0.142 (skewed) & 0.000 (equal) & -0.027 (skewed) & -0.183 (skewed) \\
C15A & -0.144 (skewed) & +0.061 (planar) & -0.043 (skewed) & -0.159 (skewed) \\
\bottomrule
\end{tabular}
\end{table*}

\begin{figure*}
\centerline{
\includegraphics[width=0.3\textwidth,angle=0]{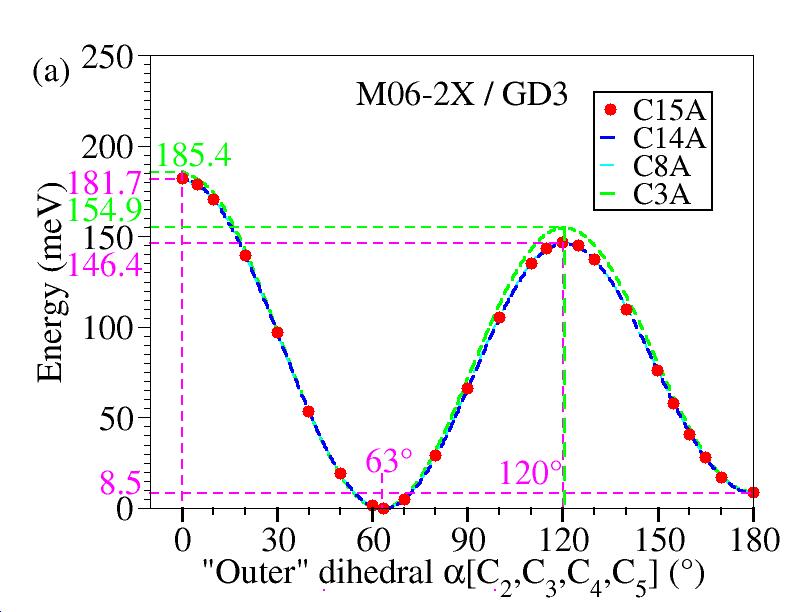}
\includegraphics[width=0.3\textwidth,angle=0]{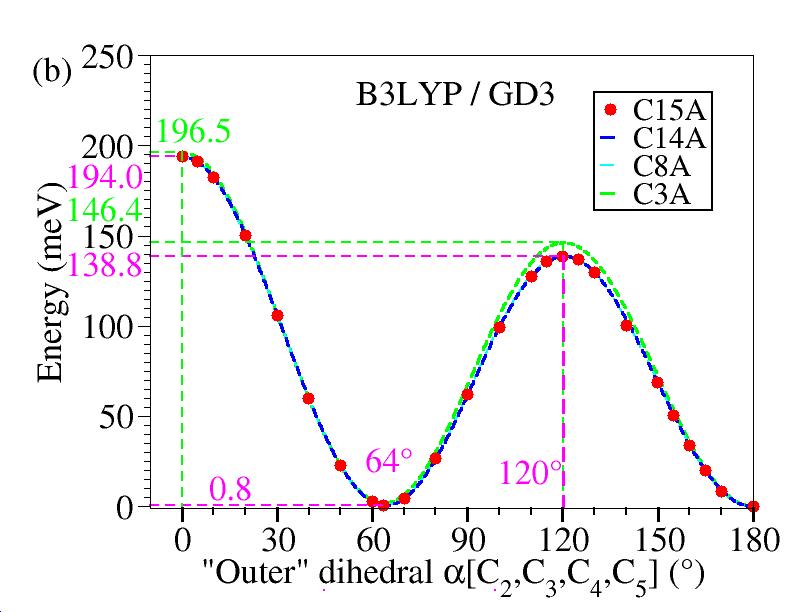}
\includegraphics[width=0.3\textwidth,angle=0]{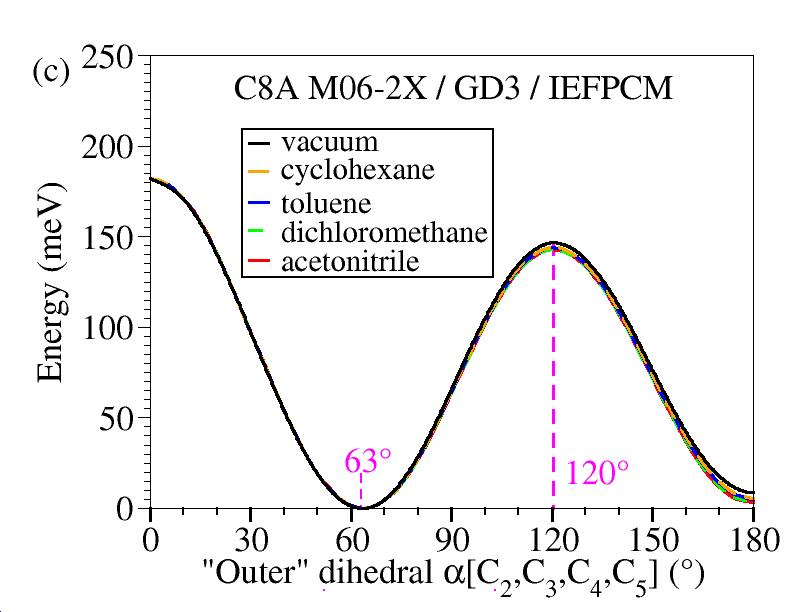}
}
\caption{Conformational energy profile for the terminal $\alpha$ dihedral angle ($\angle[\ce{C2C3C4C5}]$). Profiles shown are in vacuo for various chain lengths $n$ (a) M06-2X and (b) B3LYP, and (c) M06-2X/IEFPCM for C8A in representative solvents.
  The consistent, nearly symmetric and $\approx 150$\,meV barrier separates the near-isoenergetic planar (C$_s$) and skewed (C$_1$) minima. Solvent effects are negligible.}
\label{fig:outer-dihedral}
\end{figure*}

As illustrated in \cref{fig:outer-dihedral}a and b, the energetic ordering is highly sensitive to the functional employed, confirming the subtle and near-degenerate nature of the terminal conformational landscape.

The two states are separated by a symmetric rotational energy barrier of approximately 150\,meV (\cref{fig:outer-dihedral}). This barrier is consistent across chain lengths (n) and establishes the kinetic persistence of the individual conformers at room temperature. Furthermore, analysis using the polarizable continuum model (GAUSSIAN keyword IEFPCM \cite{g16})
confirmed that the barrier height and energetic degeneracy are robustly maintained in a range of representative solvents: non-polar (cyclohexane, toluene), weakly polar (dichloromethane), and polar (acetonitrile) (\cref{fig:outer-dihedral}c). This validated that the rotational profile is an intrinsic molecular property driven by internal electronic and steric factors.

The identification of these two near-isoenergetic conformers provides a quantum-chemical foundation for interpreting experimental observations in CnA-based systems, including recent reports of distinct transport regimes.\cite{Baldea:2026a} The C$_s$/C$_1$ duality represents an intrinsic structural property that must be accounted for in both spectroscopic analysis and device design.

\subsection*{Ab Initio Wavefunction-Based Validation of the Rotational Profile}
\label{sec:ab-initio}
It is well documented in quantum chemistry that, even for small molecules --- including many systems of astrochemical relevance \cite{Baldea:2019e,Baldea:2019g,Lam:21,Baldea:2022e} --- high-level wavefunction methods such as CCSD(T) or explicitly correlated CCSD(T)-F12 \cite{Baldea:2020e} do not always provide conformational energy differences closer to experiment than common DFT functionals or composite thermochemistry protocols (e.g., CBS-QB3, G4). In many cases, the latter approaches deliver more robust and balanced results, particularly for subtle effects dominated by dispersion and basis-set incompleteness. For this reason, we consider the near-degeneracy of the $C_s$ and $C_1$ conformers obtained from G3, G4, CBS-QB3, and CBS-4M calculations to be a reliable and robust prediction.

Nevertheless, to provide an independent wavefunction-based benchmark, we carried out relaxed potential energy scans along the terminal $\alpha$ dihedral of 1-decyne (C8A) using Dunning's cc-pVTZ basis set and the following hierarchy of methods: Hartree--Fock (HF), M{\o}ller--Plesset perturbation theory to fourth order (MP2, MP3, MP4(D), MP4(DQ), MP4(SDQ)), coupled-cluster singles and doubles (CCSD) with perturbative triples (CCSD(T) --- ``gold standard''), and quadratic configuration interaction singles and doubles (QCISD) with perturbative triples (QCISD(T)).

\begin{figure*}[!t]
\centering
\includegraphics[width=0.48\textwidth]{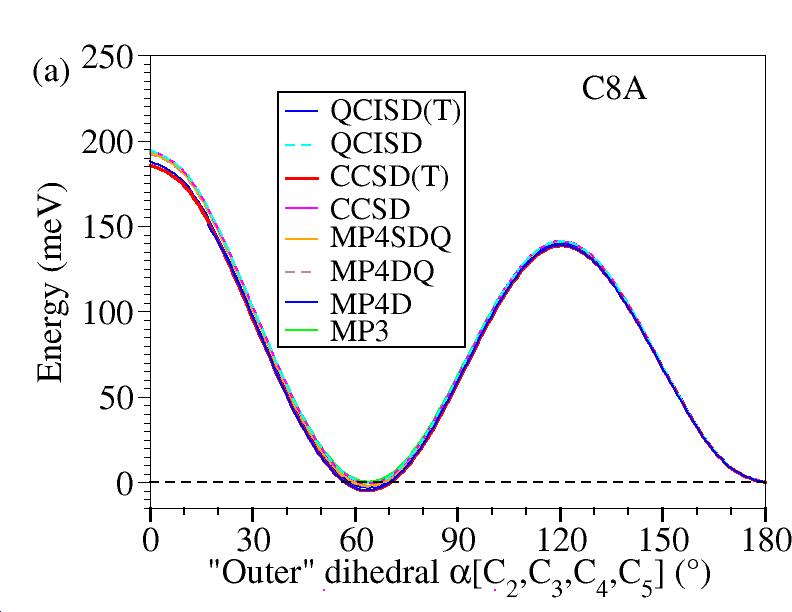}
\includegraphics[width=0.48\textwidth]{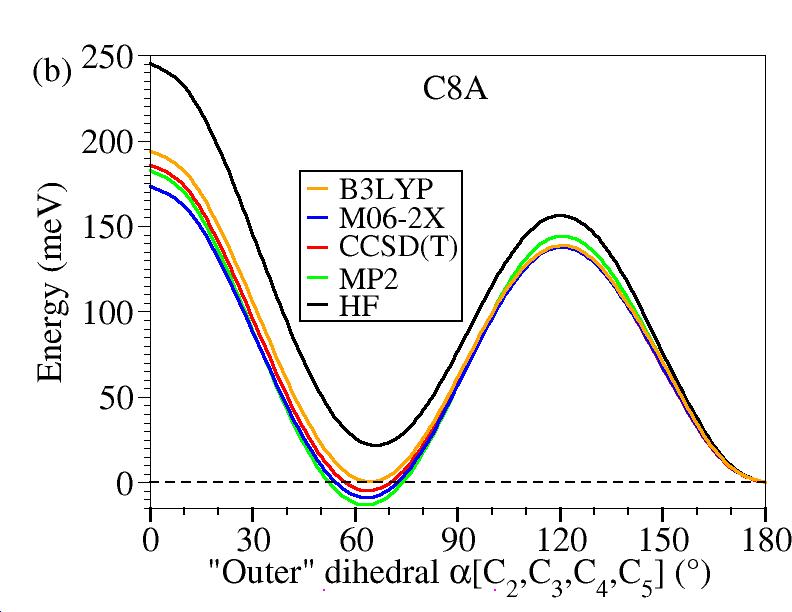}
\caption{Benchmarking the conformational energy profile of 1-decyne (C8A). The relative energy (referenced to the planar C$_s$ minimum at $\alpha = 180^\circ$) is shown as a function of the terminal $\alpha$ dihedral angle ($\angle[\ce{C2C3C4C5}]$). (a) High-level wavefunction methods computed with the cc-pVTZ basis set. The differences between the various post-Hartree–Fock methods (MP3, MP4(D), MP4(DQ), MP4(SDQ), CCSD, CCSD(T), QCISD, QCISD(T) as implemented in GAUSSIAN 16 \cite{g16}) are negligible—considerably smaller than the standard threshold for ``chemical accuracy'' ($\sim 1$ kcal/mol). Numerical values for the stationary points are provided in \Cref{tab:ab-initio}. (b) Comparison of MP2, M06-2X, and B3LYP with the CCSD(T) reference. MP2 and the two DFT functionals show reasonably small deviations from the CCSD(T) benchmark, whereas the Hartree–Fock (HF) curve deviates significantly (see \Cref{tab:ab-initio} for quantitative data).}
\label{fig:alpha-abinitio-benchmark}
\end{figure*}

The resulting potential energy profiles are collected in \cref{fig:alpha-abinitio-benchmark} and fully confirm the key qualitative and quantitative features already obtained at lower levels of theory:
\begin{itemize}
    \item two near-degenerate minima exist near $\alpha = 180^\circ$ (planar, $C_s$) and $\alpha \approx 64 - 66^\circ$ (skewed, $C_1$);
    \item the barrier maximum is located near the eclipsed geometry ($\alpha \approx 120^\circ$);
    \item the classical barrier height is consistently $\sim$135--145\,meV across correlated wavefunction methods,
      in excellent agreement with the value obtained with M06-2X and composite thermochemistry approaches.
\end{itemize}

Correlated methods from MP3 upward place the two minima within $\approx$5\,meV of each other, with most favoring the skewed structure. This confirms the near-degeneracy already observed in the composite thermochemistry protocols (G3, G4, CBS-QB3, CBS-4M).
The minor method-to-method variations---both here and in the composite methods
(cf.~\Cref{tab:conformer_stability_gibbs,tab:ab-initio} and \Cref{fig:alpha-abinitio-benchmark}a)---are well
within the standard threshold for ``chemical accuracy'' ($\sim$1\,kcal/mol) and are typical for subtle conformational energy differences
arising from the competition between hyperconjugation and dispersion-like correlation effects.

Among wavefunction-based methods, only Hartree-Fock deviates significantly, predicting a skewed $C_1$ conformer that is +21.8\,meV higher in energy than the planar form. This underscores the importance of dynamical correlation for accurately capturing the competition between hyperconjugation and dispersion-like effects.
Notably, the DFT functionals M06-2X and B3LYP perform well, deviating from the CCSD(T)/QCISD(T) benchmarks by $\approx$4.5\,meV, compared to $\sim$8\,meV for MP2
(\Cref{tab:ab-initio} and \Cref{fig:alpha-abinitio-benchmark}b).

\begin{table*}[htbp]
\centering
    \caption{Selected relative energies in meV (kcal/mol) (referenced to the planar $C_s$ minimum at $\alpha=180^\circ$) at key stationary points along the terminal $\alpha$ dihedral of 1-decyne (C8A) in vacuo. Results obtained  via relaxed potential energy scans with the indicated quantum chemical methods and cc-pVTZ basis set. The trans-eclipsed (anti-eclipsed) geometry ($\alpha\approx120^\circ$) constitutes the relevant transition state governing
  the interconversion between the planar ($C_s$) and skewed ($C_1$) minima.
  The cis-eclipsed (syn-eclipsed) conformation ($\alpha=0^\circ$) is a higher-energy secondary maximum of limited practical significance for the
    $C_s \rightleftharpoons C_1$ dynamics.}
  \label{tab:ab-initio}
\small
\begin{tabular}{l c c c c c c}
\toprule
Method          & Energy($\alpha=0^\circ$) & $\alpha_{\text{skewed}} (^\circ)$ & Energy (skewed) & $\alpha_{\text{barrier}} (^\circ)$ & Energy (barrier $\text{C}_s \to \text{C}_1$) \\
\midrule
 HF         &     245.1   (5.65)  & 66.0 &  21.8    (0.50)  & 120.1 &  156.4   (3.6)  \\
 MP2        &     182.5   (4.21)  & 63.1 & -12.9   (-0.30)  & 120.7 &  144.4   (3.3)  \\
 MP3        &     193.8   (4.47)  & 64.0 &   0.4    (0.01)  & 120.3 &  140.8   (3.2)  \\
 MP4D       &     187.9   (4.33)  & 63.8 &  -3.1   (-0.07)  & 120.4 &  139.6   (3.2)  \\
 MP4DQ      &     192.3   (4.44)  & 64.0 &  -2.4   (-0.06)  & 120.5 &  141.1   (3.2)  \\
 MP4SDQ     &     193.1   (4.45)  & 64.0 &  -1.9   (-0.04)  & 120.5 &  141.1   (3.2)  \\
 CCSD       &     194.4   (4.48)  & 64.1 &  -0.6   (-0.01)  & 120.4 &  141.2   (3.2)  \\
 CCSD(T)    &     185.9   (4.29)  & 63.7 &  -4.6   (-0.11)  & 120.4 &  138.8   (3.2)  \\
 QCISD      &     194.5   (4.49)  & 64.1 &  -0.5   (-0.01)  & 120.4 &  141.1   (3.2)  \\
 QCISD(T)   &     185.8   (4.29)  & 63.7 &  -4.5   (-0.10)  & 120.4 &  138.7   (3.2)  \\
 M06-2X/GD3 &     173.4   (4.00)  & 63.3 &  -8.5   (-0.20)  & 120.5 &  138.0   (3.1)  \\
 B3LYP/GD3  &     194.1   (4.48)  & 64.3 &   0.8    (0.02)  & 120.6 &  138.8   (3.2)  \\
\bottomrule
\end{tabular}
\end{table*}

\subsection*{Mechanism and Kinetics of Terminal $\alpha$-Dihedral Interconversion}
\label{sec:mechanism-kinetics}

In response to the need for a clearer mechanistic picture and a more quantitative kinetic interpretation of the $\alpha\approx 120^\circ$ barrier, we characterize the relevant transition state and estimate the interconversion dynamics, including a simple assessment of tunneling contributions.

The rotational energy profiles shown in \cref{fig:outer-dihedral} reveal a direct, one-dimensional torsional mechanism for interconversion between the planar $C_s$ ($\alpha=180^\circ$) and skewed $C_1$ ($\alpha\approx 63^\circ$--$66^\circ$) conformers. The kinetically relevant transition state is the trans-eclipsed (anti-eclipsed) geometry (trans-TS, $\alpha\approx 120^\circ$), where the $\ce{C3-H}$ bonds are eclipsed relative to the triple-bond axis. Frequency calculations at the M06-2X/GD3/cc-pVTZ level confirm that this stationary point is a true transition state, with a single imaginary frequency of $|\nu^\ddagger|\approx 94\,\mathrm{cm}^{-1}$ in vacuo that corresponds predominantly to the $\alpha$-torsional coordinate. As listed in \cref{tab:imag-freq-ts}, this value remains essentially invariant across a range of solvents (cyclohexane, toluene, dichloromethane, acetonitrile), mirroring the solvent robustness of the classical barrier height itself.

For completeness, the higher-energy cis-eclipsed (syn-eclip\-sed) conformation ($\alpha=0^\circ$) also exhibits a single imaginary frequency ($\approx 158$--$159\,\mathrm{cm}^{-1}$). This point, however, lies significantly above the primary barrier ($\sim 170 - 250\,\mathrm{meV}$ depending on method) and does not lie on the lowest-energy path for $C_s\rightleftharpoons C_1$ interconversion.

\begin{table}[htbp]
\centering
\caption{Imaginary frequencies (cm$^{-1}$, magnitude) of the two transition states (TS) along the $\alpha$ dihedral of 1-decyne (C8A), computed at the M06-2X/cc-pVTZ/GD3 level (IEF-PCM for solvents). The trans-eclipsed geometry (trans-TS, $\alpha\approx120^\circ$) is the relevant transition state (trans-TS) for conformer interconversion; the cis-eclipsed geometry (cis-TS, $\alpha=0^\circ$) is a higher secondary maximum.}
\label{tab:imag-freq-ts}
\small
\begin{tabular}{l c c}
\toprule
Environment     & $|\nu^\ddagger|$ ($\alpha\approx 120^\circ$) & $|\nu^\ddagger|$ ($\alpha=0^\circ$) \\
\midrule
vacuo           & 94.4 & 158.4 \\
cyclohexane     & 93.9 & 158.0 \\
toluene         & 93.9 & 158.0 \\
dichloromethane & 93.8 & 158.7 \\
acetonitrile    & 93.8 & 159.4 \\
\bottomrule
\end{tabular}
\end{table}

The classical rate constant is obtained from the Eyring equation of transition-state theory,
\begin{equation}
k(T)=\frac{k_\mathrm{B}T}{h}\exp\left(-\frac{\Delta G^\ddagger}{RT}\right),
\label{eq:eyring}
\end{equation}
where $\Delta G^\ddagger\approx 145\,\mathrm{meV}$ ($\approx 3.4\,\mathrm{kcal\,mol}^{-1}$) at 298\,K (the entropy of activation is assumed small for this rigid torsional mode). Direct application of eq.~\eqref{eq:eyring} yields
\[
k_\text{pure}\approx 2.2\times 10^{10}\,\mathrm{s}^{-1}
\qquad(\tau_\text{pure}\approx 45\,\mathrm{ps}).
\]

A more realistic description recognizes that the pure Eyring prefactor overestimates the rate for soft torsional barriers. The general form is
\begin{equation}
k=\kappa\frac{k_\mathrm{B}T}{h}\exp\left(-\frac{\Delta G^\ddagger}{RT}\right)
=A_\text{eff}\exp\left(-\frac{\Delta G^\ddagger}{RT}\right),
\label{eq:eyring-mod}
\end{equation}
in which the transmission coefficient $\kappa$ (or, equivalently, the effective prefactor $A_\text{eff}$) accounts for dynamical recrossing.

A microscopic estimate of $A_\text{eff}$ starts from the imaginary frequency of the transition state. The value $|\nu^\ddagger|\approx 94\,\mathrm{cm}^{-1}$ corresponds to an attempt frequency
\[
\nu^\ddagger = c\,|\nu^\ddagger| \approx 2.8\times 10^{12}\,\mathrm{s}^{-1}.
\]
For soft torsional barriers, dynamical recrossing typically reduces the successful-crossing probability by one to two orders of magnitude. Adopting a conservative factor of $\kappa\approx 0.03$--$0.04$ yields
\[
A_\text{eff}\approx\kappa\,\nu^\ddagger\sim 10^{11}\,\mathrm{s}^{-1}.
\]
(The same order of magnitude is routinely recovered from more elaborate treatments of internal rotation, such as phase-space or master-equation descriptions of the torsional dynamics.)

Inserting this $A_\text{eff}$ into eq.~\eqref{eq:eyring-mod} already lowers the rate from the pure value $2.2\times 10^{10}\,\mathrm{s}^{-1}$ to $\sim 3\times 10^8\,\mathrm{s}^{-1}$.
An additional, modest reduction arises from dynamic solvent friction. Polarizable-continuum calculations capture only the static solvent effect on the barrier height; the frictional damping exerted by the solvent (Kramers regime) further attenuates the rate. A conservative additional factor of $\sim 0.3$--$1$ is typical for low-viscosity organic solvents. Combining both corrections places the interconversion rate in the interval
\[
k\approx (1\text{--}4)\times 10^{8}\,\mathrm{s}^{-1},
\]
corresponding to mean conformer lifetimes of $\sim 3$--$10\,\mathrm{ns}$ at 298\,K. Given that $\kappa$ and the friction factor are order-of-magnitude estimates, and that the $\pm 10\,\mathrm{meV}$ uncertainty in $\Delta G^\ddagger$ alone shifts $k$ by a factor of $\sim 1.5$--$2$, this result should be regarded as an order-of-magnitude estimate, $k\sim 10^8$--$10^9\,\mathrm{s}^{-1}$.

Quantum tunneling corrections are negligible under ambient conditions. The simple Wigner approximation
\begin{equation}
\Gamma(T)=1+\frac{1}{24}\left(\frac{h c |\nu^\ddagger|}{k_\mathrm{B}T}\right)^2
\label{eq:wigner}
\end{equation}
with $|\nu^\ddagger|\approx 94\,\mathrm{cm}^{-1}$ yields $\Gamma\approx 1.009$ at 298\,K (a mere $\sim 0.9\,\%$ enhancement). Even the larger imaginary frequency of the cis-TS ($\sim 158\,\mathrm{cm}^{-1}$) produces only $\Gamma\approx 1.025$. These factors remain insignificant compared with the overall uncertainty arising from the modest variation of the barrier height across methods ($\sim 135 - 156\,\mathrm{meV}$). Consequently, quantum tunneling plays no meaningful role in the interconversion kinetics at room temperature.

The resulting timescale is sufficiently fast for rapid equilibration in the gas phase or in solution on typical laboratory timescales, yet slow enough relative to ultrafast spectroscopic probes (fs--ps) to allow the two conformers to coexist as a persistent $\approx 50\%:50\%$ ensemble. At the same time, the high barrier enables kinetic trapping during low-temperature synthesis, flash deposition, or device fabrication protocols,\cite{Baldea:2026a} providing a microscopic basis for the observed dual conductance regimes in CnA-based molecular junctions.

These kinetic characteristics---rapid equilibration on laboratory timescales combined with high-barrier persistence---underpin the unique behavior of the alkyne terminus as a ``Quantum Trigger Finger'' (see below).

\subsubsection*{Steric Relief at the $\alpha$-Carbon}
\begin{figure*}
\centerline{
\includegraphics[width=0.3\textwidth,angle=0]{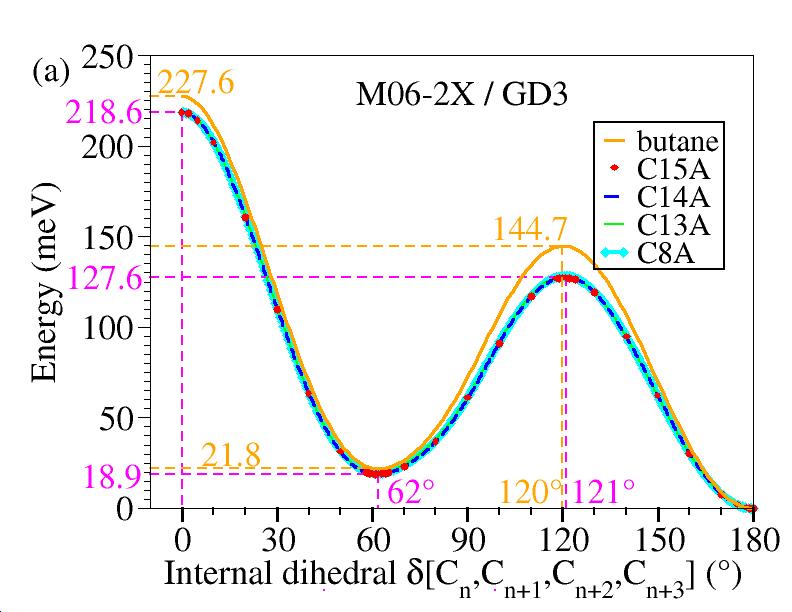}
\includegraphics[width=0.3\textwidth,angle=0]{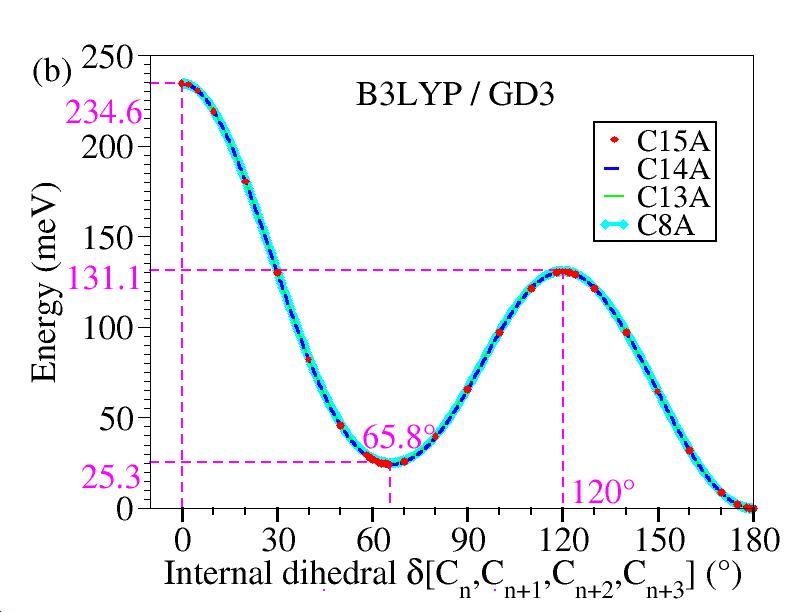}
\includegraphics[width=0.3\textwidth,angle=0]{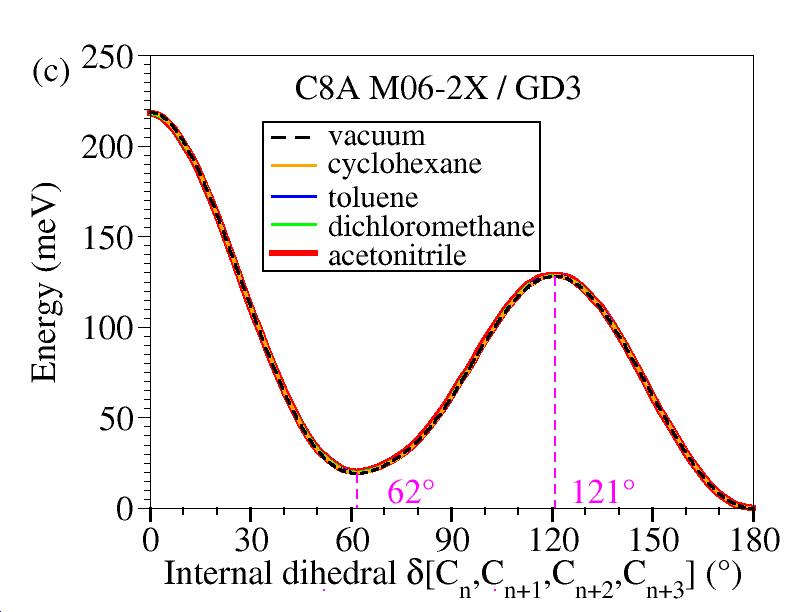}
}
\caption{Conformational energy profile for the internal $\delta$ dihedral angle ($\angle[\ce{C_{k}C_{k+1}C_{k+2}C_{k+3}}]$) along the alkyl backbone. Profiles shown are in vacuo for various chain lengths $n$ (a) M06-2X and (b) B3LYP, and (c) M06-2X/IEFPCM for C8A in solvents. The profile shows a thermodynamic preference for the $\mathrm{anti}$ (planar) state ($\approx 20$\,meV) and an asymmetric barrier ($\approx 110$\,meV vs. $\approx 130$\,meV),
  significantly lower than the symmetric $\alpha$-barrier (\cref{fig:outer-dihedral}).}
\label{fig:internal-dihedral}
\end{figure*}

The core structural difference lies in the nature of the rotating bonds. While the internal $\delta$-rotation (\Cref{fig:internal-dihedral})
involves an $\mathrm{sp}^3-\mathrm{sp}^3$ $\ce{CH2-CH2}$ bond, where the steric penalty ($\approx 20$\,meV) arises from the clash between the four hydrogen atoms on the two $\delta$ carbons, the terminal $\alpha$-rotation is fundamentally an $\mathrm{sp}-\mathrm{sp}^3$ rotation ($\mathrm{C}2-\mathrm{C}3$). The steric environment is entirely different because the $\mathrm{sp}$-hybridized carbon ($\mathrm{C}2$) is not bonded to any hydrogen atoms that could participate in a local gauche clash. This dramatic structural change provides
significant steric relief, effectively suppressing the traditional gauche penalty. With the steric cost eliminated, the stability of the non-planar C$_1$ state is governed by favorable electronic effects, specifically $\pi_{\mathrm{C}\equiv\mathrm{C}} \rightarrow \sigma^*_{\mathrm{C-H}}$ hyperconjugation, allowing it to achieve energy comparable to the planar C$_s$ state.

\subsubsection*{Validation via Fluorination}
To further probe the mechanistic origin of the $\mathrm{C}_s/\mathrm{C}_1$ near-degeneracy, the fluorinated analog,
2,2-difluoro-n-oct-1-yne (F2-C8A, \ce{H-C\equiv C-CF2-(CH2)6-CH3}), was studied
(\cref{tab:metrics-C8A_F2-C8A_m062x_w_disp}).
\begin{table*}[ht]
\small
\caption{Comparison of conformational metrics for 1-decyne (C8A = \ce{H-C\bond{3}C-(CH2)7-CH3}) and fluorinated decyne (F2-C8A = \ce{H-C\bond{3}C-CF2-(CH2)6-CH3})
  computed at the M06-2X/GD3 level of theory.\cite{g16}
\ce{X3} is the H or F atom closest to the center (M) of the triple bond \ce{C1\bond{3}C2}.}
\centering
\begin{tabular}{rrccccrccc}
\toprule
Mol.   & Conformer  & r(C1,C2) & r(C2,C3) &   r(C3,C4)   & r(C1,C10) &   d(X3,M)    & $\angle \ce{X3MC2}$  & $\alpha$ \\
\midrule
C8A    &    planar  &   1.1975 &   1.4599 &   1.5314     &     11.1065    &       2.6302       &      22.514            & 179.997  \\
F2-C8A &    planar  &   1.1942 &   1.4714 &   1.5134     &     11.0756    &       2.8271       &      26.169            & 179.999  \\
C8A    &    skewed  &   1.1977 &   1.4613 &   1.5333     &      9.6720    &       2.6239       &      22.884            &  63.256  \\
F2-C8A &    skewed  &   1.1943 &   1.4722 &   1.5120     &      9.7052    &       2.8242       &      26.489            &  60.855  \\
\bottomrule
\end{tabular}
\label{tab:metrics-C8A_F2-C8A_m062x_w_disp}
\end{table*}
Evaluation using the most reliable composite methods
(G3, G4, CBS-QB3) confirms that the near-degeneracy persists in $\mathrm{F2-C8A}$ ($\vert\Delta G \vert < 0.4\,\mathrm{kcal/mol}$), but the energetic ordering becomes slightly more pronouncedly skewed (C$_1$ favored) compared to C8A
(\cref{tab:C8A_F2-C8A-rotamer-stability-kcal-compound}).
\begin{table*}
  \caption{Conformational stabilities for 1-decyne (C8A = \ce{H-C\bond{3}C-(CH2)7-CH3}) and fluorinated decyne (F2-C8A = \ce{H-C\bond{3}C-CF2-(CH2)6-CH3}) using
    different compound chemistry models.\cite{g16} Differences in Gibbs free energy ($\Delta G$ = $G_{\mathrm{torsioned}} - G_{\mathrm{planar}}$) are given in kcal/mol.
    The negative values indicate the torsioned (skewed) rotamer is more stable.}
\centering
\begin{tabular}{l c c c}
\toprule
Method & $\Delta G$ C8A & $\Delta G$ F2-C8A \\
\midrule
G3      & -0.073  & -0.344  \\
G4      & -0.033  & -0.178  \\
CBS-QB3 & -0.068  & -0.315  \\
CBS-4M  & -0.158  & -0.113  \\
\bottomrule
\end{tabular}
\label{tab:C8A_F2-C8A-rotamer-stability-kcal-compound}
\end{table*}

This confirms the equilibrium is exquisitely sensitive to the electronic environment, as expected for a
$\Delta G \approx 0$ system. The substitution provides three key structural and electronic validations:
\begin{itemize}
    \item Steric Test Passed: Despite the larger size of the fluorine atom, the near-zero $\Delta G$ is maintained, confirming the $\mathrm{sp}$ hybridization fundamentally nullifies the steric cost of the non-planar conformation.
    \item Electronic Sensitivity Confirmed: The $\mathrm{C}2-\mathrm{C}3$ single bond lengthens upon fluorination (e.g., $1.4613\,\mathrm{\AA}$ in C8A to $1.4722\,\mathrm{\AA}$ in $\mathrm{F2-C8A}$, cf.~\cref{tab:metrics-C8A_F2-C8A_m062x_w_disp}).
      This lengthening indicates the powerful inductive effect of the $\ce{CF2}$ group strongly influences the $\sigma$-framework.
    \item Compensatory Electronic Balance: While the hyperconjugative stabilization of C$_1$ is weakened by the $\ce{C-F}$ substitution
      (distance $\mathrm{d}(\mathrm{F3},\mathrm{M})$ larger than $\mathrm{d}(\mathrm{H3},\mathrm{M})$,
      cf.~\cref{tab:metrics-C8A_F2-C8A_m062x_w_disp}),
      the overall $\Delta G$ remains near zero. This outcome reveals a critical electronic balance: the loss of $\mathrm{C-H}\dots\pi$ stabilization is effectively
      compensated by the strong inductive withdrawal of the $\ce{CF2}$ group. The $\mathrm{C}_s/\mathrm{C}_1$ balance is thus confirmed to be the result of competing electronic forces that are highly tunable by substitution.
\end{itemize}
\clearpage

\subsection*{Kinetic Contrast: Quantum ``Trigger Finger'' vs. ``Stiff Joint''}

The CnA system's kinetic profile is defined by the stark contrast between the terminal $\alpha$ rotation and the internal $\mathrm{sp}^3-\mathrm{sp}^3$ $\delta$ rotation. The comparison is best described by the analogy of a binary, kinetically locked switch versus a conventionally biased joint.

\subsubsection*{The $\mathbf{\alpha}$ Dihedral: The Kinetic Toggle Switch ("Trigger Finger")}
\label{sec:trigger-finger}

The behavior of the $\alpha$ dihedral ($\mathrm{C\equiv C-C-C}$) in n-alk-1-ynes can be described as a ``Trigger Finger'' (\emph{tenosynovitis stenosans}): a medical condition where a tendon catches on its protective sheath, causing the finger to lock abruptly in a bent position. This mechanism captures the essential kinetic nature of the $\alpha$-dihedral locking a binary state. The switch's function is defined by:

\begin{itemize}
    \item The States (Near-Degeneracy): The minimal energy difference ($\vert\Delta E\vert \leq 0.2\,\mathrm{kcal/mol}$) makes the planar (C$_s$) and staggered (C$_1$) states practically isoenergetic. This binary energy profile is the essence of the switch.
    \item The Detent ($\approx 150$\,meV Barrier): The high $\approx 150$\,meV rotational barrier acts as the molecular kinetic lock (the ``detent''). Since this energy significantly exceeds thermal energy ($\sim 25$\,meV at room temperature), interconversion is suppressed, enforcing an abrupt, switch-like behavior rather than continuous rotation.
\end{itemize}
The resulting nearly symmetric rotational energy barrier defines the chemical and kinetic behavior of the alkyne terminus, leading to the key feature of an $\approx 50\%:\,50\%$ ($\mathrm{C}_s : \mathrm{C}_1$) equilibrium mixture in gas or solution.

\subsubsection*{The $\mathbf{\delta}$ Dihedral: The Stiff Joint}

In contrast, the internal $\mathrm{sp}^3-\mathrm{sp}^3$ $\delta$ dihedral is analogous to a Stiff Joint in the human body (like a knee or elbow) that has a clear, built-in energetic preference for the extended position:

\begin{itemize}
    \item The Preferred State (Built-in Energy Penalty): The $\mathrm{anti}$ ($\delta = 180^\circ$) state is the global minimum. The $\mathrm{gauche}$ ($\delta \approx 62^\circ$) state is intrinsically less stable by $\approx 20$\,meV due to the steric clash between the four $\ce{CH2}$ hydrogen atoms.
    \item The Asymmetric Resistance: The resulting rotational barrier is asymmetric ($\approx 110$\,meV for $\mathrm{gauche} \to \mathrm{anti}$ vs. $\approx 130$\,meV for $\mathrm{anti} \to \mathrm{gauche}$). This energy difference reflects the inherent, thermodynamic preference for the $\mathrm{anti}$ state, confirming it is a conventional alkyl chain rotamer that is always biased toward linearity, resulting in a typical $\approx 80\%$ ($\mathrm{anti}$) versus $\approx 20\%$ ($\mathrm{gauche}$) distribution at room temperature.
\end{itemize}

\subsection*{Implications: Ensemble Analysis and Conformational Control}
The persistent $\mathrm{C}_s/\mathrm{C}_1$ co-existence is an intrinsic, kinetically accessible feature of the acetylenic anchor that mandates a dual approach for its application.

\begin{itemize}
\item Spectroscopic Necessity and Ensemble Average: The stable equilibrium mixture is $\approx 50\%:\,50\%$. Consequently, any measurement reflecting the state of the ensemble (e.g., standard gas- and solution-state spectroscopic data) must be interpreted as the ensemble average of these two distinct conformers. This necessitates that data analysis explicitly attempt to separate and identify the contributions from both C$_s$ and C$_1$.

\item Kinetic Enrichment and Molecular Junctions: The slow $\mathrm{C}_s \leftrightharpoons \mathrm{C}_1$ interconversion rate permits significant kinetic trapping. By utilizing an appropriate synthesis pathway and preserving the product at low temperatures, one can relatively facilely enrich the mixture, leading to an intentional imbalance in the $\mathrm{C}_s/\mathrm{C}_1$ ratio. This enriched conformational imbalance can be preserved when the molecule is integrated into a device, such as in the fabrication of molecular junctions or self-assembled monolayers, allowing for targeted studies where the impact of a non-$50\%:50\%$ distribution (the setting of the ``Trigger Finger'') can be analyzed.
\end{itemize}

\clearpage

\clearpage

\section*{Conclusion}

The present study represents a clear-cut example of theory's predictive power driving subsequent experimental discovery.
First made public as a preprint \cite{baldea_rotamers_CnA_chemrxiv} (ref.~8 in \cite{Baldea:2026a}),
the quantum-chemical prediction of two nearly isoenergetic C$_s$/C$_1$ conformers
was the key theoretical insight that preceded, guided, and was validated by
the recent experimental observation of dual conductance regimes in CnA-based molecular junctions \cite{Baldea:2026a}.

The conformational bistability revealed in this paper positions $n$-alk-1-ynes as \textit{smart molecules}
with intrinsic binary switching capability at the anchor interface.

The insights gained in this work into the origin and stabilization of the non-planar alkyne terminus are essential for interpreting spectroscopic data,
rationalizing the reactivity of alkynes, and contributing to their improved functionality as anchoring groups in molecular electronic devices.
We establish the terminal \ce{C2-C3} bond as a discretely switchable,
kinetically locked element, elevating the n-alk-1-yne anchor to a sophisticated structural motif.

By elucidating the quantum-mechanical origin of the C$_s$/C$_1$ near-degeneracy and its kinetic persistence, this work provides the theoretical foundation needed to rationalize and ultimately control conformation-dependent properties in al\-kyne-terminated molecular systems.

\section*{Acknowledgements}
I gratefully acknowledge computational
support by the state of Baden-W\"urttemberg through bwHPC and the German Research Foundation
through Grant Nos.\ INST 40/575-1, 35/1597-1, and 35/1134-1 (JUSTUS 2, bwUniCluster 2/3, and bwForCluster/MLS\&WISO/HELIX 2).

\section*{Conflict of Interest}

No conflict of interest to declare.

\section*{Ethics Statement}

No animal or human experiments were involved in this study.

\section*{Data Available Statement}

Data available on reasonable request from the author.

\begin{shaded}
\noindent\textsf{\textbf{Keywords:} \keywords}
\end{shaded}
\setlength{\bibsep}{0.0cm}

\clearpage

\section*{Table of Contents}

\noindent\rule{11cm}{2pt}
\begin{minipage}{11cm}
\includegraphics[width=11cm]{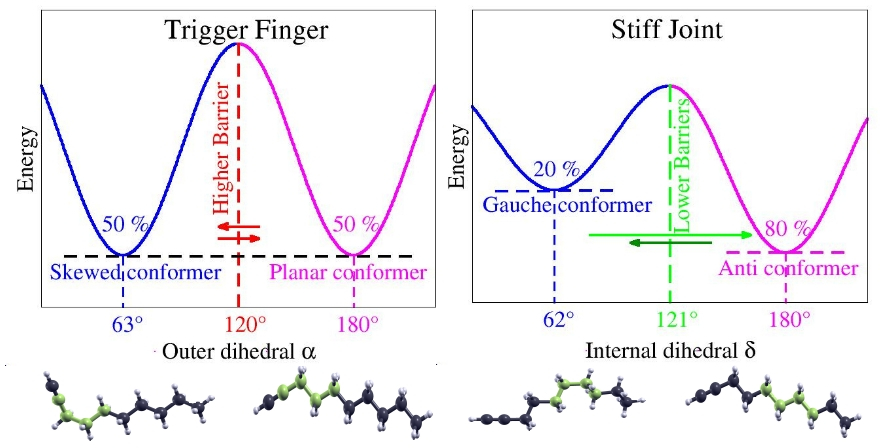}
\end{minipage}
\begin{minipage}{11cm}
  \large\textsf{
    Quantum chemical analysis reveals the unique function of n-alk-1-yne terminus as a Quantum ``Trigger Finger,''
    exhibiting two near-isoenergetic conformers
    (C$_s$ and C$_1$) locked by a symmetric high barrier. This contrasts with the lower asymmetric barrier of the
    conventional asymmetric ``Stiff Joint'' alkyl chain.
    The resulting $\approx$ 50\%:50\%
    ensemble is key for spectroscopy,
    yet permits kinetic trapping for predictable molecular junction fabrication.}
\end{minipage}
\noindent\rule{11cm}{2pt}

\end{document}